\documentclass[
    aps,prl,twocolumn,superscriptaddress
]{revtex4-2}

\usepackage{graphicx}
\usepackage[hidelinks]{hyperref}
\usepackage{bm}
\usepackage{amsmath}
\usepackage{bbold}
\usepackage{xcolor}
\usepackage{float}
\usepackage{chemformula}
\usepackage{booktabs}
\usepackage{zref-savepos}

\usepackage[normalem]{ulem} 

\begin{document}

\title{Compass Correlations in a Square-Lattice Magnet}

\renewcommand*{\thefootnote}{\arabic{footnote}}

\author{Otkur Omar}
\affiliation{Department of Physics, University of Science and Technology of China, Hefei, Anhui 230026, China}

\author{Matthias D. Frontzek}
\affiliation{Neutron Scattering Division, Oak Ridge National Laboratory, Oak Ridge, TN 37831, USA}

\author{Gabriele Sala}
\affiliation{J-PARC Center, Japan Atomic Energy Agency, Tokai, Ibaraki, Japan.}

\author{Junying Shen}
\affiliation{Institute of High Energy Physics, Chinese Academy of Sciences (CAS), Beijing 100049, China}
\affiliation{Spallation Neutron Source Science Center, Dongguan 523803, China}
\affiliation{Guangdong Provincial Key Laboratory of Extreme Conditions, Dongguan, 523803, China}

\author{Xinmin Wang}
\affiliation{Institute of High Energy Physics, Chinese Academy of Sciences (CAS), Beijing 100049, China}
\affiliation{Spallation Neutron Source Science Center, Dongguan 523803, China}
\affiliation{Guangdong Provincial Key Laboratory of Extreme Conditions, Dongguan, 523803, China}

\author{Ze Hu}
\affiliation{Institute of High Energy Physics, Chinese Academy of Sciences (CAS), Beijing 100049, China}
\affiliation{Spallation Neutron Source Science Center, Dongguan 523803, China}
\affiliation{Guangdong Provincial Key Laboratory of Extreme Conditions, Dongguan, 523803, China}

\author{Zhuang Xu}
\affiliation{Institute of High Energy Physics, Chinese Academy of Sciences (CAS), Beijing 100049, China}
\affiliation{Spallation Neutron Source Science Center, Dongguan 523803, China}
\affiliation{Guangdong Provincial Key Laboratory of Extreme Conditions, Dongguan, 523803, China}

\author{Zecong Qin}
\affiliation{Institute of High Energy Physics, Chinese Academy of Sciences (CAS), Beijing 100049, China}
\affiliation{Spallation Neutron Source Science Center, Dongguan 523803, China}
\affiliation{Guangdong Provincial Key Laboratory of Extreme Conditions, Dongguan, 523803, China}

\author{Xin Tong}
\affiliation{Institute of High Energy Physics, Chinese Academy of Sciences (CAS), Beijing 100049, China}
\affiliation{Spallation Neutron Source Science Center, Dongguan 523803, China}
\affiliation{Guangdong Provincial Key Laboratory of Extreme Conditions, Dongguan, 523803, China}

\author{Joe Paddison}
\affiliation{Materials Science \& Technology Division, Oak Ridge National Laboratory, Oak Ridge, TN 37831, USA}

\author{Max Hirschberger}
\affiliation{Department of Applied Physics, The University of Tokyo, Tokyo 113-8656, Japan}

\author{Shinichiro Seki}
\affiliation{Department of Applied Physics, The University of Tokyo, Tokyo 113-8656, Japan}

\author{Masaaki Matsuda}
\affiliation{Neutron Scattering Division, Oak Ridge National Laboratory, Oak Ridge, TN 37831, USA}

\author{Matthew B. Stone}
\affiliation{Neutron Scattering Division, Oak Ridge National Laboratory, Oak Ridge, TN 37831, USA}

\author{Andrew D. Christianson}
\affiliation{Materials Science \& Technology Division, Oak Ridge National Laboratory, Oak Ridge, TN 37831, USA}

\author{Nguyen D. Khanh}
\email[]{khanh@ap.t.u-tokyo.ac.jp}
\affiliation{Department of Applied Physics, The University of Tokyo, Tokyo 113-8656, Japan}
\affiliation{RIKEN Center for Emergent Matter Science (CEMS), Wako, 351-0198, Japan}
\affiliation{Institute of Multidisciplinary Research for Advanced Materials, Tohoku University, Sendai, Miyagi 980-8577, Japan}

\author{Shang Gao}
\email[]{sgao@ustc.edu.cn}
\affiliation{Department of Physics, University of Science and Technology of China, Hefei, Anhui 230026, China}
\affiliation{Neutron Scattering Division, Oak Ridge National Laboratory, Oak Ridge, TN 37831, USA}
\affiliation{Materials Science \& Technology Division, Oak Ridge National Laboratory, Oak Ridge, TN 37831, USA}


\begin{abstract}
The square-lattice compass model is a paradigm of bond-dependent frustration, yet experimental evidence for its characteristic spin correlations has remained elusive. Here we identify compass correlations in the square-lattice intermetallic ErRu$_2$Si$_2$ using single-crystal neutron diffuse scattering. Comparative measurements on the isostructural compounds NdRu$_2$Si$_2$ and ErRu$_2$Si$_2$ reveal a pronounced contrast between nearly  uniform diffuse rings in the former and strong intensity modulation between Brillouin zones in the latter. Through the magnetic neutron polarization factor, this modulation reveals compass correlations that couple moment orientation to the correlation wavevector. Crystal-field spectroscopy identifies four preferred in-plane orientations for the Er$^{3+}$ moments. A minimal compass-Heisenberg model incorporating this anisotropy quantitatively reproduces the diffuse scattering and accounts for the double-$Q$ vortex-antivortex ground state. Fits with the compass couplings fixed at values set by the dipolar energy scale retain good agreement with the data, supporting a substantial dipolar contribution to the bond-dependent interactions. Our results establish a square-lattice platform for compass physics and suggest that dipolar interactions, acting together with local crystal-field anisotropy, can shape directional magnetic correlations alongside conduction-electron-mediated exchange.
\end{abstract}

 \maketitle
\textit{Introduction.---}
While the isotropic Heisenberg model has long served as a cornerstone of magnetism, the compass model, characterized by bond-dependent anisotropic interactions, has opened new avenues in condensed matter physics and quantum information technology~\cite{Nussinov_C_2015}. One prominent example is the compass model on a honeycomb lattice, most notably the exactly solvable Kitaev model~\cite{KITAEV_F_2003}. This system is celebrated for the emergence of topological order, where fractionalized Majorana fermions coexist with static $\mathbb{Z}_2$ gauge fields~\cite{KITAEV_A_2006}. The realization of such exotic phases has firmly established the compass interaction as a fundamental mechanism for driving unconventional correlations~\cite{jackeli_mott_2009, Nussinov_C_2015,takagi_concept_2019, banerjee_neutron_2017, do_majorana_2017, evered_probing_2025}.

Despite progress on the honeycomb lattice, establishing compass correlations in square-lattice magnets remains an experimental challenge.  The square-lattice compass model exemplifies how bond-dependent interactions generate frustration on a geometrically unfrustrated lattice, supporting directional nematic order selected by fluctuations~\cite{Nussinov_C_2015,Mishra_D_2004,Wenzel_M_2008}. Its symmetries have also motivated quantum error-correcting codes and proposals for protected qubits~\cite{Bacon_PRA_2006,Doucot_PRB_2005, Milman_PRL_2007}. In models incorporating additional interactions, compass couplings can help stabilize topological spin textures, including meron and skyrmion crystals~\cite{Wang_M_2021,Hirschberger_Nanometric_2021}. Candidate materials include square-lattice iridates such as Ba$_2$IrO$_4$ and Sr$_2$IrO$_4$, in which compass interactions have been proposed~\cite{Katukuri_PRX_2014,Zhang_PRB_2016}. However, their magnon dispersions are largely governed by isotropic exchange, complicating the identification of weaker bond-dependent contributions~\cite{Kim_PRL_2012, Clancy_PRB_2023}. Establishing clear compass correlations would therefore provide an experimental basis for testing how bond-dependent interactions organize magnetic fluctuations and select the ordered state.

\begin{figure*}[t]
\centering
\includegraphics[width=0.95\textwidth]{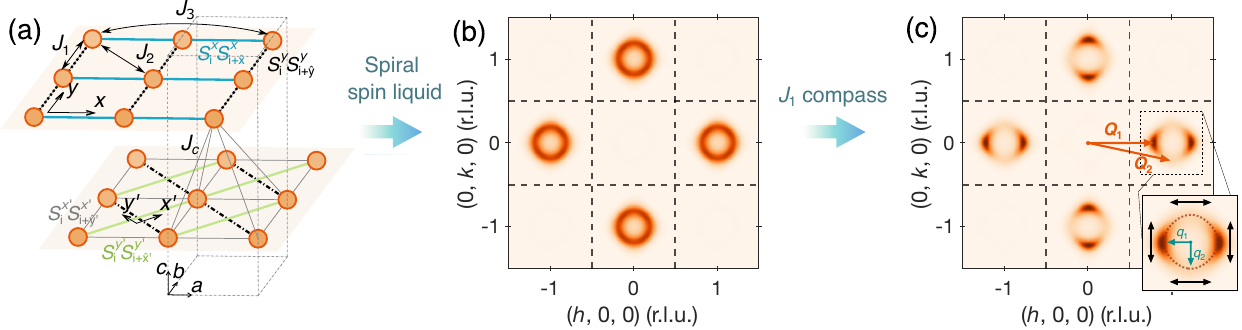}
\caption{Lattice geometry and compass correlations. (a) AB-stacked square lattices of rare-earth $R^{3+}$ ions (orange spheres) in $R$Ru$_2$Si$_2$. Double-headed arrows indicate the first-, second-, and third-neighbor Heisenberg exchange paths, $J_1$, $J_2$, and $J_3$. And the interlayer coupling $J_c$ is also indicated. Colored bonds illustrate compass couplings on the $J_1$ (upper layer) and $J_2$ (lower layer) bonds. (b) Calculated diffuse scattering for a spiral spin liquid with $J_1=-1$, $J_2=0.4$, $J_3=0.2$, and $J_c=0.1$~meV at $T=6$~K. The continuous rings reflect the degeneracy of the spiral states. (c) Calculated pattern with an additional compass coupling $J_1^{\rm C}=-0.2$~meV. The magnetic polarization factor modulates the rings between Brillouin zones. $\mathbf{Q}_{1,2}$ and $\mathbf{q}_{1,2}$ denote momentum transfers and propagation vectors; black arrows indicate preferred moment orientations.}
\label{fig:intro}

\end{figure*}

To address this challenge, we compare two isostructural magnets, $R$Ru$_2$Si$_2$ ($R$ = Nd, Er), whose rare-earth moments occupy square lattices but develop contrasting magnetic ground states~\cite{Shigeoka_magnetic_1993,Garnier2000,Sato2001}. In \ch{NdRu2Si2}, single-$Q$ order with propagation vector $\mathbf{q} = (0.13, 0.13, 0)$ emerges below the transition temperature of $T{_\mathrm{N}} = 23$~K~\cite{Shigeoka_magnetic_1993}, whereas \ch{ErRu2Si2} adopts double-$Q$ order with $\mathbf{q}_1 = (0.2, 0, 0)$ and $\mathbf{q}_2 = (0, 0.2, 0)$ below $T{_\mathrm{N}} = 6$~K~\cite{Garnier2000, Sato2001}. We probe the interactions underlying this contrast through comparative neutron scattering in their correlated paramagnetic phases. By exploiting the sensitivity of the magnetic polarization factor to moment orientation, we obtain direct evidence for compass correlations in \ch{ErRu2Si2}. A minimal compass-Heisenberg model incorporating the four-state clock-like anisotropy inferred from crystal-field spectroscopy quantitatively reproduces both the paramagnetic correlations and the double-$Q$ ground state in \ch{ErRu2Si2}. Energy-scale analysis and quantitative fitting further show that effective compass couplings arise when classical dipolar interactions are projected onto this four-state clock manifold. The resulting connection between local crystal-field anisotropy and dipolar coupling establishes a microscopic route to compass physics in rare-earth intermetallics.

\textit{Compass correlations probed by diffuse neutron scattering.---}
Figure~\ref{fig:intro}(a) depicts the AB-stacked square lattice formed by the rare-earth ions in $R$Ru$_2$Si$_2$, where the first-, second-, and third-neighboring bonds, denoted $J_1$,~$J_2$,~and $J_3$, are indicated by double-headed arrows in the upper layer. In the absence of compass interactions, competition between these couplings may stabilize a spiral spin liquid phase~\cite{bergman_order_2007, yao_generic_2021}, characterized by continuous diffuse scattering rings in reciprocal space resulting from macroscopic degeneracy~\cite{gao_spiral_2017, gao_spiral_2022, wan_spiral_2024, andriushin_observation_2025, gao_codimension_2026}. A representative pattern for such a phase is shown in Fig.~\ref{fig:intro}(b). However, in the presence of compass interactions, the correlated magnetic moments will exhibit a directional preference depending on the propagation vector, $\bf{q}$. For illustration, Fig.~\ref{fig:intro}(c) presents a scenario where ferromagnetic compass couplings on the $J_1$ bonds, described by $\mathcal{H}_1^{\rm{C}} = J^{\rm{C}}_1\sum_i (S_i^xS_{i+\hat{x}}^x + S_i^yS_{i+\hat{y}}^y )$ as indicated in Fig.~\ref{fig:intro}(a), enhance spin correlations perpendicular to $\bf{q}$. For magnetic neutron scattering, the cross section, $S(\mathbf{Q})$, depends only on the spin components that are perpendicular to the momentum transfer $\bf{Q}$,
$ S(\mathbf{Q}) \propto |M_{\perp}(\mathbf{Q})|^{2}.$
Therefore, neutron scattering intensity in the region around $\bf{Q} \parallel \bf{q}$ and $\bf{Q} \perp \bf{q}$ will be enhanced and suppressed, respectively, leading to a Brillouin zone-dependent modulation of the scattering intensity as illustrated in Fig.~\ref{fig:intro}(c).

\begin{figure*}[t]
\centering
\includegraphics[width=0.8\textwidth]{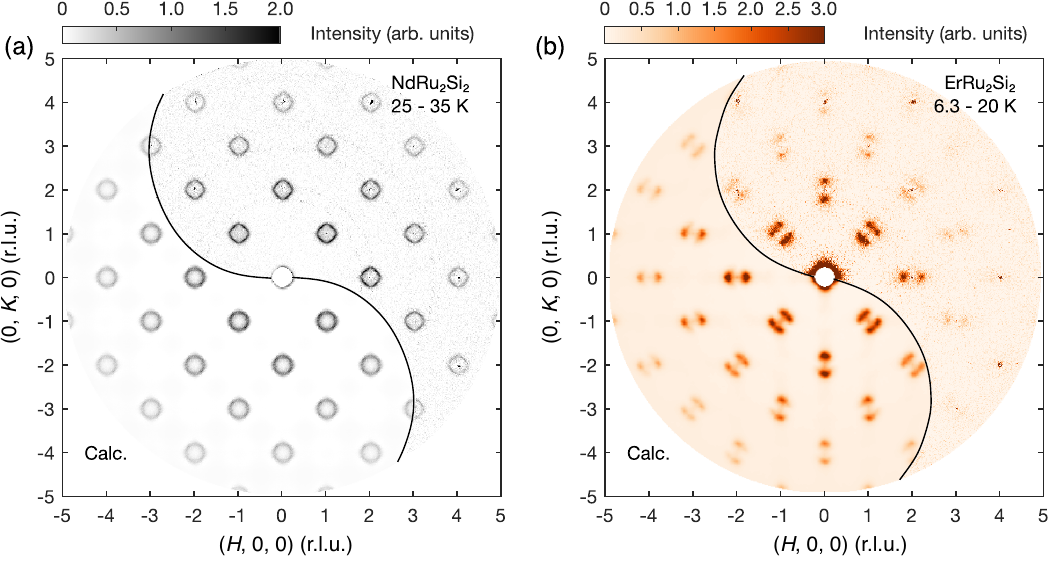}
\caption{Diffuse neutron scattering signatures of compass correlations in the $(H,K,0)$ plane. (a) NdRu$_2$Si$_2$: the upper-right region shows WAND$^2$ data at $T=25$~K with the 35~K background subtracted, and the lower-left region shows the fitted Ising model with $J_1=-1.58(3)$, $J_2=0.69(1)$, $J_3=0.42(1)$, and $J_c=-0.187(5)$~meV. (b) ErRu$_2$Si$_2$: the upper-right region shows WAND$^2$ data at $T=6.3$~K with the 20~K background subtracted, and the lower-left region shows the fitted compass-Heisenberg model. Black curves separate experiment and calculation. }
\label{fig:diffuse}

\end{figure*}

\begin{figure}[t]
\centering
\includegraphics[width=0.9\columnwidth]{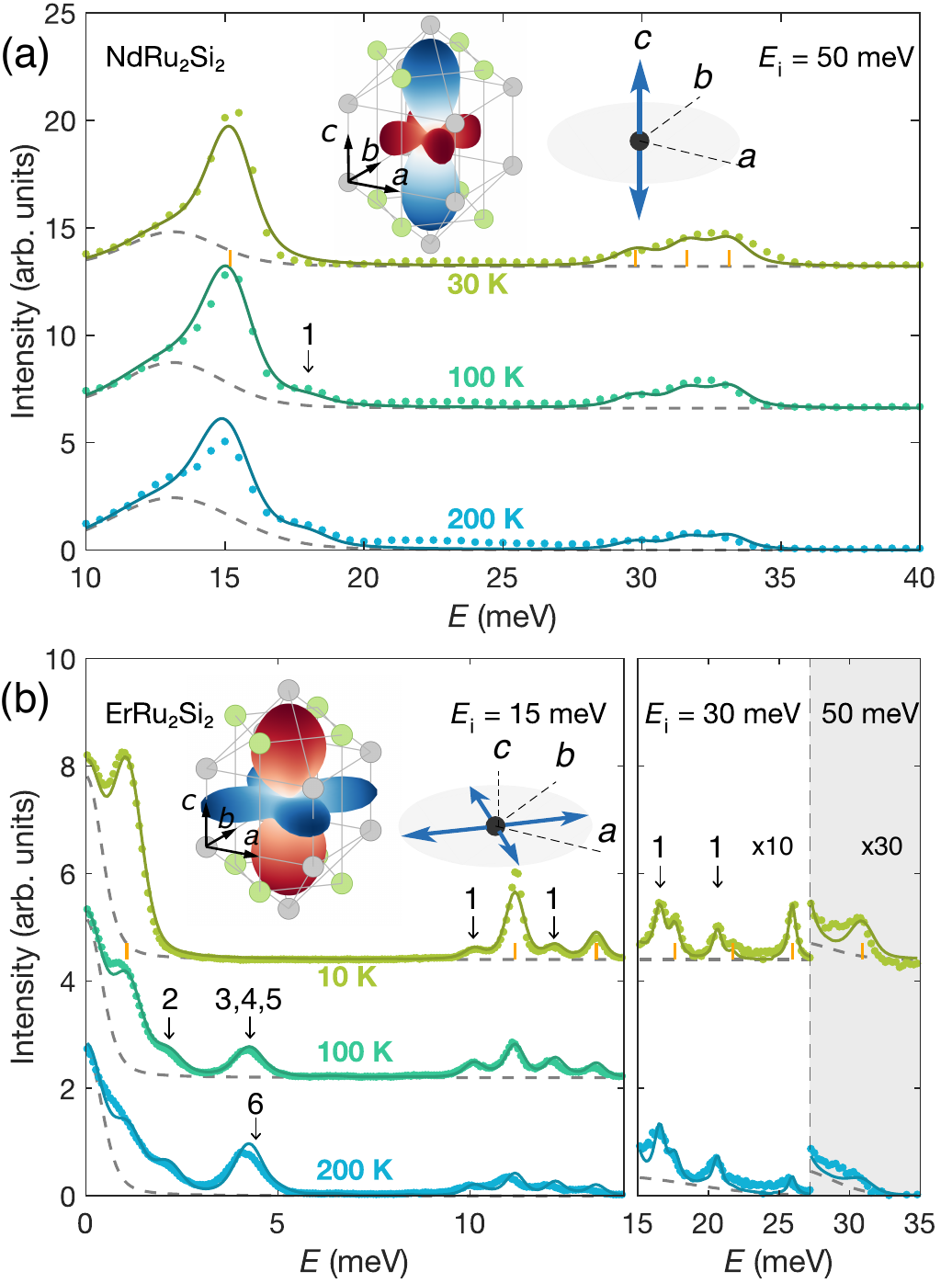}
\caption{Crystal-field spectra and single-ion anisotropy. (a) NdRu$_2$Si$_2$ spectra measured on ARCS with $E_i=50$~meV and integrated over $Q=2.3$--$3.7$~\AA$^{-1}$. (b) ErRu$_2$Si$_2$ spectra measured on HD with $E_i=15$, 30, and 50~meV and integrated over $Q=2.0$--$2.3$, 2.6--3.2, and 2.2--2.8~\AA$^{-1}$, respectively. Circles denote data, solid curves CEF fits, and gray dashed curves fitted backgrounds. Typical error bars are smaller than the markers. Yellow bars mark CEF energies. Numbered arrows identify transitions from thermally populated excited levels by their initial-level index. Traces are offset by 6.6 units in (a) and 2.2 units in (b); the $E_i=30$ and 50~meV spectra in (b) are scaled by 10 and 30. Insets show calculated anisotropy-energy surfaces (red, white, and blue denote positive, zero, and negative energies) and the $c$-axis Ising and in-plane clock easy directions.}
\label{fig:cef}
\end{figure}

Figures~\ref{fig:diffuse}(a) and (b) compare the diffuse neutron scattering maps in the $(H,K,0)$ plane for NdRu$_{2}$Si$_{2}$ and ErRu$_{2}$Si$_{2}$, respectively. Measurements were performed in the correlated paramagnetic regime for both compounds on the WAND$^2$ diffractometer at the Oak Ridge National Laboratory (ORNL), using floating-zone-grown single crystals. The 35~K (Nd) and 20~K (Er) data were subtracted as backgrounds. Experimental procedures are given in the Supplemental Material (SM)~\cite{supp}. For NdRu$_{2}$Si$_{2}$, the scattering intensity forms continuous rings around the centers of the Brillouin zones, which is a characteristic signature of spiral spin liquids as exemplified in Fig.~\ref{fig:intro}(b). In contrast, the scattering rings shown in  Fig.~\ref{fig:diffuse}(b) for ErRu$_{2}$Si$_{2}$ are all incomplete and follow a distinct zone dependence. For example, for the scattering ring around $(1,1,0)$, only the top-right and bottom-left quarter arcs are visible, while the visibility is reversed around $(-1,1,0)$.  Following the illustration of Fig.~\ref{fig:intro}(c), this intensity modulation evidences that the correlated spin components tend to be perpendicular to the propagation vectors, thus revealing the emergence of compass correlations in ErRu$_{2}$Si$_{2}$.

\textit{Effective compass-Heisenberg model.---}
In rare-earth magnets, the single-ion states define the local degrees of freedom from which an effective spin Hamiltonian is constructed. The crystal electric field (CEF) splits the $(2J+1)$-dimensional ground-state multiplet of each $R^{3+}$ ion and determines its magnetic anisotropy, where $J$ denotes the total angular momentum. We therefore use inelastic neutron scattering to characterize the CEF excitations in \ch{NdRu2Si2} and \ch{ErRu2Si2}. 

Figures~\ref{fig:cef}(a) and (b) show the temperature-dependent integrated inelastic neutron spectra of NdRu$_2$Si$_2$ and ErRu$_2$Si$_2$, respectively. The measurements were performed using the ARCS spectrometer at ORNL for NdRu$_2$Si$_2$ and the HD spectrometer~\cite{luo_design_2023} at the China Spallation Neutron Source (CSNS) for ErRu$_2$Si$_2$. The corresponding $S(\bm{Q},\omega)$ maps are presented in the SM~\cite{supp}. The broad wavevector coverage of the NdRu$_2$Si$_2$ measurements allows us to identify a phonon mode near 13~meV, which is represented by a Gaussian contribution in the integrated spectra [gray dashed line in Fig.~\ref{fig:cef}(a)]. For ErRu$_2$Si$_2$, the gray dashed lines in Fig.~\ref{fig:cef}(b) represent the fitted elastic-peak tail in the $E_i=15$~meV spectra and the broad background contributions in the $E_i=30$ and 50~meV spectra. Fits to the temperature-dependent integrated spectra in Figs.~\ref{fig:cef}(a) and (b) yield the CEF parameters and ground-state doublet wave functions summarized in Table~\ref{tab:cef} in the Appendix.

The resulting anisotropy-energy profiles shown in the insets of Figs.~\ref{fig:cef}(a) and (b)~\cite{supp} reveal contrasting single-ion anisotropies: Nd$^{3+}$ exhibits Ising anisotropy along the $c$ axis, whereas Er$^{3+}$ exhibits four-state clock-like anisotropy with four symmetry-equivalent easy directions along the in-plane $\langle110\rangle$ directions. Both results agree with previous magnetization measurements~\cite{shigeoka_magnetic_1990,Sato2001}. The Ising character of the Nd$^{3+}$ moments also clarifies why complete diffuse rings are visible in Fig.~\ref{fig:diffuse}(a). Moments along $c$ remain perpendicular to every scattering vector $\bm{Q}$ in the $(H,K,0)$ plane, giving a constant magnetic polarization factor that leaves all portions of the rings observable. An Ising spin model reproduces the observed diffuse pattern~\cite{supp}, as shown in the lower-left portion of Fig.~\ref{fig:diffuse}(a).

The four-state clock anisotropy of the Er$^{3+}$ moments arises from virtual transitions involving the first excited CEF doublet, located approximately 1.07~meV above the ground state~\cite{supp}. To describe the diffuse scattering measured at $T=6.3$~K, where excited CEF states are appreciably populated, we construct an effective compass-Heisenberg model incorporating this single-ion anisotropy. The analysis in the SM shows that compass couplings are required on both the $J_1$ and $J_2$ bonds~\cite{supp}, giving
\begin{equation}
\begin{aligned}
\mathcal{H}={}&J_1^{\rm C}\sum_i
\left(S_i^xS_{i+\hat{x}}^x+S_i^yS_{i+\hat{y}}^y\right)\\
+&J_2^{\rm C}\sum_i
\left(S_i^{x'}S_{i+\hat{x}'}^{x'}+S_i^{y'}S_{i+\hat{y}'}^{y'}\right)\\
+&\sum_{n\in{1,2,3,c}}J_n^{\rm H}
\sum_{\langle ij\rangle_n}\mathbf{S}_i\cdot\mathbf{S}_j .
\end{aligned}
\label{eq:model}
\end{equation}
The $J_2$ compass term is expressed in the rotated $x'$–$y'$ frame, whose axes lie along the square-lattice diagonals, as illustrated in the lower layer of Fig.~\ref{fig:intro}(a). The final term describes isotropic Heisenberg interactions, with $\langle ij\rangle_n$ denoting bonds connecting first, second and third neighbours within the $ab$ plane ($n=1,2,3$) and the shortest interlayer bonds ($n=c$), as indicated in Fig.~\ref{fig:intro}(a).

\begin{figure}[t!]
\centering
\includegraphics[width=\columnwidth]{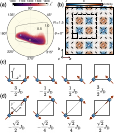}
\caption{Vortex-antivortex ground state and dipolar contributions in ErRu$_2$Si$_2$. (a) Polar $\chi^2$ map with $R\cos\theta=J_1^{\rm H}/|J_1^{\rm C}+J_1^{\rm H}|$ and $R\sin\theta=J_2^{\rm C}/|J_1^{\rm C}+J_1^{\rm H}|$. The dashed boundary encloses the double-$Q$ regime; the star and blue circle mark the optimal compass-Heisenberg and dipolar-Heisenberg models. (b) Double-$Q$ ground-state spin configuration from classical Monte Carlo simulations. Black and white arrows denote moments in the $c=0$ and $c=1/2$ planes. The dashed square is the magnetic unit cell, and the background color denotes local vorticity in the $c=0$ plane. (c), (d) Dipolar energies of representative four-state clock configurations on nearest- and next-nearest-neighbor bonds, respectively, in units of $J_{\rm D}=\mu_0\mu^2/(4\pi a^3)$.}
\label{fig:order}
\end{figure}

We fit the diffuse neutron scattering data by least squares using static spin correlations obtained from classical Monte Carlo simulations of Eq.~\eqref{eq:model} on a $20\times20\times10$ supercell with periodic boundary conditions. Calculated patterns are differences between the target and background temperatures, as in the experiment. More numerical details are presented in the SM~\cite{supp}. Initial unconstrained fits reveal a strong linear correlation between the interaction parameters, $J_2^{xx}=-0.536J_1^{yy}$, where $J_2^{xx}=J_2^{\rm H}+\frac{1}{2}J_2^{\rm C}$ and $J_1^{yy}=J_1^{\rm H}$~\cite{supp}. We therefore impose this relation in the subsequent analysis. Figure~\ref{fig:order}(a) shows the resulting $\chi^2$ landscape, with $J_3^{\rm H}$ and $J_c^{\rm H}$ fixed at their optimal values of 0.08 and $-0.07$~meV, respectively. The polar coordinates are defined by $R\cos\theta=J_1^{\rm H}/|J_1^{\rm C}+J_1^{\rm H}|$ and $R\sin\theta=J_2^{\rm C}/|J_1^{\rm C}+J_1^{\rm H}|$. A well-defined valley of low $\chi^2$ contains the optimal parameter set, $\{J_1^{\rm H},J_1^{\rm C},J_2^{\rm H},J_2^{\rm C}\}=\{-0.02(2),-0.56(2),0.15(1),-0.27(1)\}$~meV, marked by the star in Fig.~\ref{fig:order}(a). The optimized model closely reproduces the diffuse scattering throughout the measured reciprocal-space volume, including the $(H,K,0)$ plane [Fig.~\ref{fig:diffuse}(b)] and cuts at different $L$ values (Fig.~S3 of the SM~\cite{supp}). This agreement provides strong support for compass correlations in ErRu$_2$Si$_2$.

To test whether the compass-Heisenberg model also captures the magnetic order in ErRu$_2$Si$_2$, we examine its ground state using parallel-tempering Monte Carlo simulations with the parameters obtained from the diffuse-scattering fits~\cite{supp}. The low-$\chi^2$ region in Fig.~\ref{fig:order}(a) lies within the double-$Q$ ground-state regime indicated by the dashed boundary~\cite{supp}. The corresponding spin configuration forms a topological vortex–antivortex lattice [Fig.~\ref{fig:order}(b)], consistent with the magnetic structure previously determined by neutron diffraction~\cite{Sato2001,Garnier2000}. The agreement shows that a single set of interaction parameters accounts for both the paramagnetic correlations and the ordered spin texture, providing further support for the effective compass-Heisenberg description of ErRu$_2$Si$_2$.

\textit{Dipolar contributions to the compass interactions.---}
We next examine whether classical dipolar interactions, acting within the four-state clock manifold, can account for the effective compass couplings in ErRu$_2$Si$_2$. For unit-length spins, the dipolar interaction between sites $i$ and $j$ takes the form \begin{equation} E_{ij}^{\mathrm{dip}} = D_{ij}\left[ \mathbf{S}_i\cdot\mathbf{S}_j -3(\mathbf{S}_i\cdot\hat{\mathbf{r}}_{ij}) (\mathbf{S}_j\cdot\hat{\mathbf{r}}_{ij}) \right], \end{equation} where $D_{ij}=\mu_0\mu^2/(4\pi r_{ij}^3)$ and $\hat{\mathbf{r}}_{ij}$ denotes the bond direction. This expression consists of an isotropic contribution and a ferromagnetic compass term that couples the moment components along the bond. The local clock anisotropy restricts the moments to four in-plane orientations, allowing this bond dependence to be expressed directly within the effective compass-Heisenberg model.

Figures~\ref{fig:order}(c) and (d) illustrate the dipolar energies of representative clock-state configurations on the $J_1$ and $J_2$ bonds, respectively. Taking $\mu=9\,\mu_{\mathrm{B}}$ and the nearest-neighbor distance $a=4.138$~\AA\ gives a dipolar energy scale of $J_{\mathrm{D}}=\mu_0\mu^2/(4\pi a^3)\approx0.0614$~meV. On the $J_1$ bonds, the two lowest-energy configurations both have ferromagnetically aligned bond-parallel components, consistent with the negative $J_1^{\mathrm{C}}$ obtained from the diffuse scattering fit. On the diagonal $J_2$ bonds, dipolar interactions favor ferromagnetic alignment along the bond and antiferromagnetic alignment perpendicular to it. Configurations with mutually perpendicular moments have zero dipolar energy and are omitted for clarity. These preferences are captured by an isotropic contribution $D_{ij}$ together with a compass coupling $-3D_{ij}$, yielding \begin{equation} \begin{aligned} J_1^{\mathrm{C}} &= -3J_{\mathrm{D}} = -0.184~\mathrm{meV},\\ J_2^{\mathrm{C}} &= -\frac{3J_{\mathrm{D}}}{2\sqrt{2}} = -0.065~\mathrm{meV}. \end{aligned} \end{equation}

To assess whether these dipolar contributions can account for the measured correlations, we fit the diffuse scattering data using a dipolar-Heisenberg model that retains dipolar interactions on the $J_1$ and $J_2$ bonds. The compass couplings are fixed at the values above, and only the additional isotropic exchanges are varied. The best fit gives $J_1^{\mathrm{H}}=-0.34(4)$, $J_2^{\mathrm{H}}=0.23(3)$, $J_3^{\mathrm{H}}=0.087(5)$, and $J_c^{\mathrm{H}}=-0.091(4)$~meV, where the reported $J_1^{\mathrm{H}}$ and $J_2^{\mathrm{H}}$ include the isotropic dipolar contributions. These parameters are consistent with the correlation between $J_2^{xx}$ and $J_1^{yy}$ identified in the unrestricted compass-Heisenberg fit. Their position in the $\theta$--$R$ plane, marked by the blue circle in Fig.~\ref{fig:order}(a), falls within the same low-$\chi^2$ valley. The calculated diffuse patterns closely reproduce the experimental data~\cite{supp}, with a $\chi^2$ value only $6\%$ above the unrestricted optimum. Thus, compass couplings fixed by the dipolar energy scale provide a good description of the observed correlations, supporting a substantial dipolar contribution to the effective bond-dependent interactions.

Additional bond-dependent exchange may arise from spin-orbit-mediated anisotropic RKKY interactions, which can contain a symmetric anisotropic component~\cite{hermenau_stabilizing_2019}. A pseudo-dipolar contribution of the form $K_{ij}^{\mathrm{pd}}[3(\mathbf{S}_i\cdot\hat{\mathbf{r}}_{ij})(\mathbf{S}_j\cdot\hat{\mathbf{r}}_{ij})-\mathbf{S}_i\cdot\mathbf{S}_j]$ would also contribute to the effective compass couplings. Establishing its magnitude and sign in ErRu$_2$Si$_2$ requires a microscopic treatment of the electronic structure and exchange processes. While the present analysis does not uniquely determine the microscopic origin, it identifies classical dipolar interactions acting within the local clock manifold as a quantitatively plausible route to compass correlations in this square-lattice magnet.

\textit{Discussion---}
The correlations in ErRu$_2$Si$_2$ connect square-lattice compass physics to an experimentally accessible metallic magnet. Their characteristic modulation across Brillouin zones reveals a coupling between moment orientation and correlation wavevector, resolved through the polarization factor of magnetic neutron scattering. A minimal compass-Heisenberg model incorporating the crystal-field-derived clock anisotropy accounts for both these paramagnetic correlations and the double-$Q$ vortex-antivortex ground state. This connection between short-range correlations and magnetic order identifies bond-dependent interactions as a common organizing principle across the two regimes.

The possible dipolar origin of these interactions is particularly significant in an intermetallic, where conduction-electron-mediated RKKY exchange provides the conventional starting point for understanding magnetic correlations. Compass couplings fixed by the dipolar energy scale retain a close quantitative description of the diffuse scattering, suggesting that dipolar interactions can shape its directional character even in the presence of substantial exchange. The emerging picture is one in which local crystal fields constrain the available moment orientations, while the bond dependence of dipolar interactions helps select how those moments correlate. This offers a microscopic route to compass correlations distinct from the superexchange mechanisms commonly invoked in spin-orbit-entangled Mott insulators~\cite{jackeli_mott_2009,Nussinov_C_2015,takagi_concept_2019}.

The metallic setting raises a further question: how do compass correlations affect the electrons that mediate part of the magnetic interaction? Conduction electrons also scatter from the correlated moments, creating a feedback between electronic motion and the directional structure of magnetic fluctuations. Determining whether this structure produces identifiable transport signatures would connect compass physics to electronic behavior, potentially even above the onset of magnetic order. Fields or strain could help test this connection by modifying the correlations and their electronic response. This opens the prospect of using metallic transport to probe compass correlations and exploring how bond-dependent magnetism shapes transport in frustrated metals.

\begin{acknowledgments}
We acknowledge helpful discussions with Zhijie Fan. This work was funded by the National Key R\&D Program of China under the Grant No. 2024YFA1613100 and the National Natural Science Foundation of China (NSFC) under the Grant No. 12374152. Work at the UTokyo was partly supported by JSPS KAKENHI under Grant No. 25K00956. J. Shen would like to acknowledge the support from Joint Funds of the National Natural Science Foundation of China (Grant No. U2430209). X. Tong would like to acknowledge the support from the National Science Fund for Distinguished Young Scholars (No. 12425512). M.H. was supported by JSPS KAKENHI Grants Nos. JP25K24552, JP26H00644, JP26H01290, JST FOREST Grant No. JPMJFR2238, and by the Deutsche Forschungsgemeinschaft (DFG, German Research Foundation) via Transregio TRR 360 - 492547816. This work was supported by the U.S. Department of Energy, Office of Science, Basic Energy Sciences, Materials Sciences and Engineering Division. This research used resources at the Spallation Neutron Source (SNS) and the High Flux Isotope Reactor (HFIR), both of which are DOE Office of Science User Facilities operated by the Oak Ridge National Laboratory (ORNL). The beam time was allocated to ARCS on proposal number IPTS-29924, to WAND$^2$ on proposal numbers IPTS-28421 and IPTS-33220, and to HB1 on proposal number IPTS-28926. 
\end{acknowledgments}

\nocite{BLAISE1994171,CFjl,HAMADA1995907,abernathy_design_2012,arnold_mantid_2014,azuah_dave_2009,buessen_spinmc_2021,classical_Zhang_2021,dahlbom_Sunnyjl_2025,faramarzi_Marine_2020a,frontzek_wand_2018,zhong_MOMPA_2021}
\bibliography{ErRu2Si2_v2.bib}

\clearpage
\appendix*
\onecolumngrid
\noindent\textit{Appendix: Crystal-field parameters and wave functions.---}
Table~\ref{tab:cef} summarizes the CEF parameters and low-energy wave functions obtained from the fits in Fig.~\ref{fig:cef}.

\begin{table*}[h!]
\caption{
    Fitted CEF parameters, calculated energy levels of the excited states, and ground-state wavefunctions $|\psi_0^\pm\rangle$ for \ch{NdRu2Si2} and \ch{ErRu2Si2}. For \ch{ErRu2Si2}, the wavefunction $|\psi_1^\pm\rangle$ of the low-lying first excited state is also shown.
}  
\label{tab:cef}
\begin{ruledtabular}
\begin{tabular*}{\textwidth}{@{\extracolsep{\fill}} l ccccc}
& $B_2^0$ (meV) & $B_4^0$ & $B_4^4$ & $B_6^0$ & $B_6^4$ \\
\colrule 
\ch{NdRu2Si2} & -4.534$\times10^{-1}$ & -1.825$\times10^{-3}$ & -1.951$\times10^{-2}$ & 5.545$\times10^{-5}$ & -1.061$\times10^{-3}$ \\
\ch{ErRu2Si2} & 1.579$\times 10^{-1}$ & -1.727$\times 10^{-4}$ & 7.901$\times 10^{-4}$ & 9.544$\times 10^{-7}$ & -3.820$\times 10^{-5}$ \\
\end{tabular*}
\begin{tabular*}{\columnwidth}{@{\extracolsep{\fill}} l c c c c c c c c}
  & $E_1$ (meV) & $E_2$ & $E_3$ & $E_4$ & $E_5$ & $E_6$ & $E_7$ \\
\colrule
\ch{NdRu2Si2} &  15.19 & 29.75 & 31.63 & 33.14 & -- & -- & -- \\
\ch{ErRu2Si2} &  1.07 & 11.17 & 13.29 & 17.63 & 21.70 & 25.93 & 30.89 \\
\end{tabular*}
\begin{tabular*}{\columnwidth}{@{\extracolsep{\fill}} l c c}
\ch{NdRu2Si2}& $|\psi_0^\pm\rangle$ : $\tiny 0.914|\mp\frac{9}{2}\rangle + 0.389 |\mp\frac{1}{2}\rangle+0.113|\pm\frac{7}{2}\rangle $ \\
\ch{ErRu2Si2}& $|\psi_0^\pm\rangle$ : $\tiny \pm0.113|\mp\frac{13}{2}\rangle \mp 0.0998|\pm\frac{11}{2}\rangle \pm0.591|\mp\frac{5}{2}\rangle \mp0.792|\pm\frac{3}{2}\rangle $\\
&$|\psi_1^\pm\rangle$: $\tiny -0.0465|\mp\frac{15}{2}\rangle -0.311|\mp\frac{7}{2}\rangle +0.9488|\pm\frac{1}{2}\rangle -0.0335|\pm\frac{9}{2}\rangle $

\end{tabular*}
\end{ruledtabular}
\end{table*}

\end{document}